\documentclass[sigconf, nonacm, pdfa]{acmart}

\usepackage{graphicx} % Required for inserting images
\usepackage{tikz}
\usepackage{listings}
\usepackage{multirow}
\usepackage{multicol}
\usepackage{booktabs}
\usepackage{url} 
\usepackage{xspace}
\usepackage{balance}
\usepackage{float}
\usepackage{makecell}
\usepackage{enumitem}
\usepackage{array}
\usepackage{utfsym} 
\usepackage{textcomp}
\usepackage{CJKutf8}
\usepackage{tabularx}

\usepackage{svg}
\usepackage{caption}
\usepackage{subcaption}
\usepackage{circledtext}
\usepackage{adjustbox}

\usepackage[a-2b]{pdfx}

\newcommand\redout{\bgroup\markoverwith
	{\textcolor{red}{\rule[.45ex]{1.5pt}{1.pt}}}\ULon}

 \newcommand\blueout{\bgroup\markoverwith
	{\textcolor{blue}{\rule[.45ex]{1.5pt}{1.pt}}}\ULon}

\definecolor{light-gray}{gray}{0.80}
\definecolor{ForestGreen}{RGB}{63,147,88}

\makeatletter
\newenvironment{btHighlight}[1][]
{\begingroup\tikzset{bt@Highlight@par/.style={#1}}\begin{lrbox}{\@tempboxa}}
	{\end{lrbox}\bt@HL@box[bt@Highlight@par]{\@tempboxa}\endgroup}

\newcommand\btHL[1][]{%
	\begin{btHighlight}[#1]\bgroup\aftergroup\bt@HL@endenv%
	}
	\def\bt@HL@endenv{%
	\end{btHighlight}%   
	\egroup
}
\newcommand{\bt@HL@box}[2][]{%
	\tikz[#1]{%
		\pgfpathrectangle{\pgfpoint{1pt}{0pt}}{\pgfpoint{\wd #2}{\ht #2}}%
		\pgfusepath{use as bounding box}%
		\node[anchor=base west, fill=orange!25,outer sep=.5pt,inner xsep=0.5pt, inner ysep=0.15pt, rounded corners=1pt, minimum height=\ht\strutbox-.1pt,#1]{\raisebox{.01pt}{\strut}\strut\usebox{#2}};
	}%
}
\makeatother

\lstdefinestyle{codestyle}{
	frame=none,
	commentstyle=\color{gray},
        columns=fixed, 
        keywordstyle=\color{black}\bfseries,
	numberstyle=\tiny\color{gray},
	stringstyle=\color{purple},
        basicstyle=\fontsize{7.9}{7.9}\ttfamily,
	breakatwhitespace=false,         
	breaklines=false,                 
	captionpos=t,                    
	keepspaces=true,     
	numbers=none,                    
	numbersep=4pt,                  
	showspaces=false,                
	showstringspaces=false,
	showtabs=false,                  
	tabsize=2,
	language=Sql,
	escapechar=|`,
        moredelim=**[is][{\btHL[fill=red!30]}]{@}{@},
        moredelim=**[is][{\btHL[fill=blue!30]}]{^}{^},
        morekeywords={with},
}

\usepackage{soul}
\usetikzlibrary{calc}

\makeatletter
\newif\if@anonymize

\@anonymizefalse  % Uncomment to show text

\if@anonymize
  \newcommand{\highlight@DoHighlight}{
    \fill [outer sep = -15pt, inner sep = 0pt, color=black]
          ($(begin highlight)+(0,8pt)$) rectangle ($(end highlight)+(0,-3pt)$) ;
  }

  \newcommand{\highlight@BeginHighlight}{
    \coordinate (begin highlight) at (0,0) ;
  }

  \newcommand{\highlight@EndHighlight}{
    \coordinate (end highlight) at (0,0) ;
  }

  \newdimen\highlight@previous
  \newdimen\highlight@current
  \newlength{\item@width}

  \DeclareRobustCommand*\anonymize{%
    \SOUL@setup
    \def\SOUL@preamble{%
      \begin{tikzpicture}[overlay, remember picture]
        \highlight@BeginHighlight
        \highlight@EndHighlight
      \end{tikzpicture}%
    }%
    \def\SOUL@postamble{%
      \begin{tikzpicture}[overlay, remember picture]
        \highlight@EndHighlight
        \highlight@DoHighlight
      \end{tikzpicture}%
    }%
    \def\SOUL@everyhyphen{%
      \discretionary{%
        \SOUL@setkern\SOUL@hyphkern
        \SOUL@sethyphenchar
        \tikz[overlay, remember picture] \highlight@EndHighlight ;%
      }{%
      }{%
        \SOUL@setkern\SOUL@charkern
      }%
    }%
    \def\SOUL@everyexhyphen##1{%
      \SOUL@setkern\SOUL@hyphkern
      \settowidth{\item@width}{##1}%
      \makebox[\item@width]{}%
      \discretionary{%
        \tikz[overlay, remember picture] \highlight@EndHighlight ;%
      }{%
      }{%
        \SOUL@setkern\SOUL@charkern
      }%
    }%
    \def\SOUL@everysyllable{%
      \begin{tikzpicture}[overlay, remember picture]
        \path let \p0 = (begin highlight), \p1 = (0,0) in \pgfextra
          \global\highlight@previous=\y0
          \global\highlight@current =\y1
        \endpgfextra (0,0) ;
        \ifdim\highlight@current < \highlight@previous
          \highlight@DoHighlight
          \highlight@BeginHighlight
        \fi
      \end{tikzpicture}%
      \settowidth{\item@width}{\the\SOUL@syllable}%
      \makebox[\item@width]{}%
      \tikz[overlay, remember picture] \highlight@EndHighlight ;%
    }%
    \SOUL@
  }
\else
  \newcommand{\anonymize}[1]{#1}
\fi
\makeatother

\usepackage[dvipsnames,table]{xcolor}
\usepackage[linesnumbered,ruled,vlined]{algorithm2e}
\DontPrintSemicolon

\SetKwComment{Comment}{\color{RawSienna}// }{}

\newcommand{\assign}{\leftarrow}

\SetKwProg{Function}{function}{}{}

\usepackage{pifont}

\usepackage{tcolorbox}
\tcbuselibrary{breakable} % Add this if required

\usepackage{pbalance}

\setlist[itemize]{leftmargin=*} % Removes indentation for all itemize lists

\newcommand\vldbdoi{10.14778/3849398.3849411}
\newcommand\vldbpages{5148 - 5161}
\newcommand\vldbvolume{19}
\newcommand\vldbissue{13}
\newcommand\vldbyear{2026}
\newcommand\vldbauthors{\authors}
\newcommand\vldbtitle{\shorttitle} 
\newcommand\vldbavailabilityurl{https://github.com/pixelsdb/pixels/tree/master/pixels-retina}
\newcommand\vldbpagestyle{empty} 

\begin{document}

\title{Transactional Data Change for Real-Time Analytics in Lakehouses}

%%
%% The "author" command and its associated commands are used to define the authors and their affiliations.

\author{Dongyang Geng}
\affiliation{
  \institution{Renmin University of China}
}
\email{gengdongyang@ruc.edu.cn}

\author{Haoqiong Bian}
\authornote{Haoqiong Bian is the corresponding author, and he contributed equally with Dongyang Geng in this research.}
\affiliation{%
  \institution{Renmin University of China}
}
\email{bianhaoqiong@ruc.edu.cn}

\author{Zinuo Li}
\affiliation{
  \institution{Renmin University of China}
}
\email{lizinuo@ruc.edu.cn}

\author{Haoyue Li}
\affiliation{
  \institution{Renmin University of China}
}
\email{lihaoyue03@ruc.edu.cn}

\begin{abstract}
  Real-time data analytics has become a key demand for lakehouses.
  It requires propagating data changes from the data source to the lakehouse in real time, while ensuring ACID guarantees and minimizing resource consumption and impact on analytical query performance.
  This is highly challenging for a data system according to the RUM conjecture, especially given the read-optimized immutable file storage layer in lakehouses.
  Existing lakehouses offer only git-style catalog version control on a single table, and they trade data freshness, query performance, and/or resource efficiency to achieve a barely acceptable data-change throughput.
  In this paper, we propose a novel data synchronization framework, Retina,
  that enables real-time, cost-efficient, query-performant, and fully transactional data changes in lakehouses. 
  We implement and open-source Retina.
  Evaluations show that Retina delivers 10-millisecond-level data freshness and over 3.2 million row/s linear-scalable data-change throughput with low resource consumption and query-performance overhead, significantly outperforming state-of-the-art lakehouse systems and lake-compatible HTAP databases.
\end{abstract}

\maketitle

\pagestyle{\vldbpagestyle}
\begingroup\small\noindent\raggedright\textbf{PVLDB Reference Format:}\\
\vldbauthors. \vldbtitle. PVLDB, \vldbvolume(\vldbissue): \vldbpages, \vldbyear.\\
\href{https://doi.org/\vldbdoi}{doi:\vldbdoi}
\endgroup
\begingroup
\renewcommand\thefootnote{}\footnote{\noindent
This work is licensed under the Creative Commons BY-NC-ND 4.0 International License. Visit \url{https://creativecommons.org/licenses/by-nc-nd/4.0/} to view a copy of this license. For any use beyond those covered by this license, obtain permission by emailing \href{mailto:info@vldb.org}{info@vldb.org}. Copyright is held by the owner/author(s). Publication rights licensed to the VLDB Endowment. \\
\raggedright Proceedings of the VLDB Endowment, Vol. \vldbvolume, No. \vldbissue\ %
ISSN 2150-8097. \\
\href{https://doi.org/\vldbdoi}{doi:\vldbdoi} \\
}\addtocounter{footnote}{-1}\endgroup

\ifdefempty{\vldbavailabilityurl}{https://github.com/pixelsdb/pixels/tree/master/pixels-retina}{
\vspace{.3cm}
\begingroup\small\noindent\raggedright\textbf{PVLDB Artifact Availability:}\\
The source code, data, and/or other artifacts have been made available at \url{\vldbavailabilityurl}.
\endgroup
}
\section{Introduction}\label{sec:introduction}

Modern data lakes, often called lakehouses~\cite{lakehouse}, bring the advantages of data warehouses in caching, indexing, and metadata management into data lakes, enabling more real-time and performant data analytics on cheap, scalable lake storage composed of open columnar file formats and object storage or Hadoop Distributed File System (HDFS).
Real-time analytics requires synchronizing changes to the transactional data source, including tuple inserts, deletes, and updates, to the lakehouse with low latency and high throughput.
Modern lakehouses such as Apache Iceberg~\cite{iceberg} and Paimon~\cite{paimon} employ change-data-capture (CDC) systems such as Debezium~\cite{debezium} to extract data changes from the transaction logs of the data source, and send them to the lakehouse for replay.
Such log-based CDC can capture complete, up-to-date data changes with minimal interference to online transactional processing (OLTP), eliminating the expense of extract-transform-load (ETL).

Although CDC solves the extraction issue, the key challenge remains how to promptly replay data changes in the lakehouse with transaction guarantees. Unlike conventional relational databases, the lakehouse prioritizes cost efficiency and scalability, and its storage layer is tuned for analytical queries dominated by large sequential scans and does not support in-place updates~\cite{lakehouse}.
The RUM (\underline{R}ead performance, \underline{U}pdate performance, and \underline{M}emory or storage costs) conjecture proposes that optimizing any two aspects in a data system may negatively impact the third~\cite{rum-conjecture-paper}. This implies that, given the lakehouse's tendency toward low costs and performant reads, simultaneously supporting high-performance updates is highly challenging. For transaction guarantees, a multi-version mechanism is necessary to enable concurrent data changes and queries, as data is not in-place updatable. However, conventional version-chain-based multi-version concurrency control (MVCC) protocols widely used in relational databases~\cite{heatwave-in-depth,hana-paper,hyper-paper,colibri-paper,umbra-mvcc,empirical-mvcc} are not applicable to lakehouses. Building and traversing version chains in lake storage is inefficient given the immutable file model and high random I/O latency, while keeping chains in memory incurs significant memory overhead and scalability issues.

\begin{figure}
     \centering
     \begin{subfigure}[b]{0.2\textwidth}
         \centering
         \includegraphics[width=\textwidth]{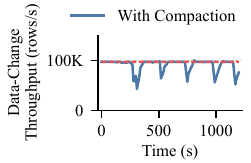}
         \caption{Data-change throughput.}
         \label{subfig:compaction-throughput}
     \end{subfigure}
     \begin{subfigure}[b]{0.2\textwidth}
         \centering
         \includegraphics[width=\textwidth]{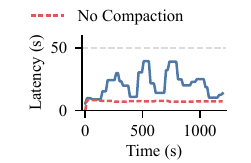}
         \caption{Checkpoint latency.}
         \label{subfig:compaction-freshness}
     \end{subfigure}
     \caption{Performance impact of compaction on data changes.}
     \Description{}
     \label{fig:compaction-overhead}
\end{figure}

Existing lakehouse systems, such as Iceberg~\cite{iceberg} and Paimon~\cite{paimon}, do not support real-time and fully transactional data changes~\cite{lake-tables-short}.
As a compromise, they implement a git-style catalog version-control mechanism~\cite {lake-tables-short}: creating a new snapshot of the table catalog via a checkpoint operation that atomically adds/removes a batch of files to/from the table.
This provides limited ACID guarantees on a single table at the file granularity.
To replay data changes, these systems use a merge-on-read (MoR) approach, which writes new tuple versions to delta files that are checkpointed in batches.
Delta files in the visible catalog snapshot are merged with the base data files to resolve visible tuple versions at query runtime.
MoR prioritizes write performance but incurs significant overhead during query execution.
Lakehouse systems such as Paimon introduce background compaction to mitigate the merge overhead~\cite{paimon-compaction}.
However, compaction is CPU- and I/O-intensive.
It causes resource contention and back pressure on data-change replay, leading to large fluctuations in replay throughput and data freshness.
As shown in Figure~\ref{fig:compaction-overhead}, we run the data-change workloads extracted from HyBench~\cite{hybench} on Paimon with and without background compaction.
We observe that during compaction, data-change throughput drops by approximately half, while checkpoint latency increases by about fourfold, severely reducing data freshness.

In this paper, we propose Retina, a data synchronization framework for lakehouses that replays
the transactional data changes at high throughput and low latency, without compromising query performance or cost efficiency.
Based on the observations that storage space is the cheapest resource and that sequential scan is no longer a major performance bottleneck in modern lakehouses,
Retina pushes down the visibility check as a vectorized filter operation into the table scan stage of queries.
Thus, each scan task performs visibility checks independently on its data split and passes only visible tuple versions downstream.
This vectorized-filter-on-read (VFoR) approach is far more efficient and scalable than MoR, and it eliminates the expensive compaction of delta and base data.
The storage space occupied by stale tuple versions is reclaimed lazily by incremental storage garbage collection jobs.

Retina replays committed write transactions from the data source as ACID mirror transactions in the lakehouse, with the same isolation level and granularity. To enable this, we propose a lightweight MVCC protocol that coordinates transaction replay and query processing using two timestamp watermarks: the high watermark (HWM), which means all mirror transactions earlier than it are replayed and visible to queries, and the low watermark (LWM), which means all analytical queries earlier than it have finished.
Each mirror transaction receives a monotonically increasing timestamp and advances HWM as it is replayed.
Each query receives $(HWM-1)$ as its snapshot timestamp and advances LWM when it finishes.
%Queries are running in the window of $[LWM, HWM)$.
This protocol isolates mirror transactions and queries into separate timestamp windows, thus avoiding write-read conflicts.
Meanwhile, the properties of the CDC stream (see Section~\ref{subsec:preliminaries-cdc}) ensure there are no conflicts among mirror transactions if they obtain timestamps in the order they appear in the stream.
Except for timestamp allocation and watermark maintenance, the protocol requires no additional state synchronization or conflict resolution.

The efficiency of VFoR benefits from a novel version-storage scheme that cooperates with the MVCC protocol. We observe that only the tuple versions deleted within $(LWM, HWM)$ might have uncertain visibility across running queries.
Therefore, we propose an in-memory data structure, \textit{tile visibility}, that determines visibility alongside the create timestamp on tuples.
It keeps only the delete timestamps of tuple versions deleted after $LWM$ and masks the tuple versions deleted earlier using bitmaps.
As $LWM$ advances, old delete timestamps gradually collapse into the bitmaps to free up memory.
Checkpoint and recovery mechanisms address durability and long-query stall issues for tile visibility.
In evaluations, tile visibility consumes only 1.56 GB (1.16 bit/tuple) of memory for HyBench SF1333 (1TB) under high data-change throughput.
Visibility calculation on tile visibility is accelerated by SIMD instructions, adding negligible overhead to query execution.

We implement and open-source Retina as the data synchronization framework in Pixels (https://github.com/pixelsdb/pixels), an efficient lakehouse engine for on-premises and cloud data analytics.

In summary, the main contributions of this paper are:

\begin{itemize}
    \item We reveal that in lakehouses, (1) catalog version control can not provide full ACID guarantees for fine-grained transactions, and (2) MoR and background compaction are the major bottlenecks hindering high-performance, cost-efficient data change and query processing.
    Hence, we propose a data synchronization framework, Retina, 
    that supports real-time transactional data changes in lakehouses without compromising query performance, scalability, or resource efficiency. (Sections~\ref{sec:preliminaries} and~\ref{sec:system-overview}).

    \item In Retina, we propose a novel version-storage scheme and the corresponding lightweight MVCC protocol to support fine-grained cross-table transactions and eliminate MoR and data compaction. It is based on the observations that (1) filtering out stale tuple versions during table scan is more efficient than MoR+compaction for query processing, and (2) the natural properties of CDC can help simplify MVCC in lakehouses. (Sections~\ref{sec:data-storage} and~\ref{sec:query-processing}).

    \item Evaluations on HyBench~\cite{hybench} and CH-benCHmark~\cite{chbenchmark} show that Retina simultaneously provides 10-ms-level data freshness, over 3.2 million row/s linear-scalable replay throughput, and transaction guarantees, without compromising query performance or cost-efficiency, significantly outperforms state-of-the-art lakehouses (Iceberg~\cite{iceberg} and Paimon~\cite{paimon}) and lake-friendly HTAP databases (TiDB~\cite{tidb-paper} and Colibri~\cite{colibri-paper}). (Section~\ref{sec:evaluation}).
\end{itemize}
\section{Preliminaries}\label{sec:preliminaries}

In this section, we introduce how data is managed in today's lakehouses and how data changes are captured by the change-data-capture (CDC) systems and synchronized to lakehouses.

\subsection{Lakehouse}\label{subsec:lakehouse}

The lakehouse architecture is proposed primarily to address data governance issues in data lakes~\cite{delta-lake-paper,lakehouse}.
Data lakes have emerged since the early 2010s.
They store data in open file formats (e.g., Parquet~\cite{parquetMainPage}) on distributed file systems (e.g., HDFS~\cite{HDFS:paper}) or cloud object storage (e.g., AWS S3~\cite{aws-s3}).
Raw data can be quickly loaded into data lakes and analyzed in situ following the schema-on-read paradigm.
This enables easy, prompt, cost-efficient, and scalable data access for various applications.
However, loose metadata management cannot ensure data consistency during ETL and record updates, leading to data swamps.

Lakehouse table formats, such as DeltaLake~\cite{delta-lake-paper}, Iceberg~\cite{iceberg}, and Hudi~\cite{hudi}, add a table catalog layer to the data lake.
Data is still stored in open file formats on lake storage, but the catalog layer provides file-level ACID guarantees for catalog and metadata operations (e.g., adding or removing a batch of files or a column).
This is achieved by a logging or snapshot mechanism in the catalog layer~\cite{lake-tables-short}.
For each catalog operation, the lakehouse performs a checkpoint operation that writes a new log entry or snapshot for the table catalog.
Thus, the analytical query can obtain a consistent view of a table's catalog.

In recent years, lakehouses have been widely adopted for analytical applications across finance, retail, and telecom et.al.
In these industries, data freshness is increasingly essential~\cite{real-time-data-lake-dremio,real-time-data-lake-aws,paimon}, and the data sources are usually transactional systems that undergo random data changes.
Real-time data analytics requires synchronizing data changes from data sources to the lakehouse in sub-seconds~\cite{real-time-data-lake-dremio}.

\subsection{Change Data Capture}\label{subsec:preliminaries-cdc}

The first step in real-time data change is to capture fresh data-change events in the data source.
Log-based Change data capture (CDC) is designed to address this problem.
It captures complete data changes from the write-ahead log or binlog of the source database in real time while introducing minimal impact on OLTP performance.
Popular log-based CDC systems include Debezium~\cite{debezium}, Kafka Connect~\cite{kafka-connect}, Flink CDC~\cite{flink-cdc}, and AWS DMS~\cite{aws-dms-cdc} etc.

\begin{figure}
     \centering
     \begin{subfigure}[b]{0.2\textwidth}
         \centering
         \includegraphics[width=\textwidth]{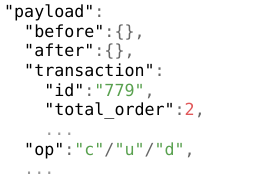}
         \caption{Tuple-change message.}
         \label{subfig:debezium-record}
     \end{subfigure}
     \begin{subfigure}[b]{0.2\textwidth}
         \centering
         \includegraphics[width=\textwidth]{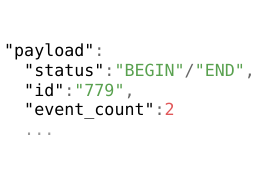}
         \caption{Transaction message.}
         \label{subfig:debezium-trans}
     \end{subfigure}
     \caption{Message structures in log-based CDC (Debezium).}
     \label{fig:debezium-message}
     \Description{}
\end{figure}

As an example, Debezium defines a source interface to connect source databases and a sink interface to connect lakehouses.
As shown in Figure~\ref{fig:debezium-message}, Debezium captures two message types from the source database.
A tuple-change message is generated for the change event (insert, update, or delete) of each tuple.
In this message, \textit{before}, \textit{after}, \textit{transaction}, and \textit{op} are the before image (tuple value before change), after image (tuple value after change), transaction information, and operation type (\textit{c} for insert, \textit{u} for update, and \textit{d} for delete), respectively.
Before is null for insert, and after is null for delete.
A transaction message is generated for each transaction begin or commit (end) event.
In this message, \textit{id} is the transaction ID, and \textit{event\_count} is the number of tuple changes in the transaction.
All messages for a transaction are captured and sent continuously upon transaction commit.
The transaction ID in the tuple-change message correlates it to the transaction, and \textit{total\_order} identifies its order within the transaction.
The messages can be passed to the lakehouse through a message queue such as Kafka~\cite{kafka-paper}.

\subsection{Data-Change Replay}\label{subsec:preliminaries-replay}

Data-change messages arriving at the lakehouse should be replayed on the base data promptly.
Unlike most OLTP and HTAP databases, which store data in in-place-updatable formats on high-performance storage media such as memory and SSDs, lakehouses store data in open columnar formats on cheap, scalable lake storage.
Open columnar formats, such as Parquet~\cite{parquetMainPage} and ORC~\cite{ORCMainPage}, are highly read-optimized.
A table stored in these formats is horizontally split into row groups.
Each row group is compressed and stored in columns.
One or more row groups are stored in one immutable file in the lake storage.
Therefore, to physically update a single row, we must read the file into memory, decompress the corresponding row group, update the column values, recompress the updated row group and file, and rewrite the updated file.
This causes significant write amplification and computational overhead.

Moreover, memory efficiency is essential for data-change replay.
Analytical queries are memory-intensive because they handle large intermediate results in memory.
Meanwhile, lakehouses prioritize cost efficiency, aiming to provide the required performance and capacity at a lower cost.
Memory is one of the most expensive resources.
It is two orders of magnitude more expensive than lake storage~\cite{fargate-pricing,s3-pricing}.
Therefore, we should minimize the memory footprint of data-change replay to free up memory for analytical queries, thus making it more affordable and applicable in lakehouses.

In addition to cost-efficiency, lakehouses prioritize high scalability and elasticity.
Scalability requires no single-point bottleneck in the system.
Elasticity requires that the query engine has no heavy states; thus, it can scale with fluctuating query load.
Therefore, concurrency control protocols that require maintaining complex data structures (e.g., in-memory version chains~\cite{heatwave-in-depth,hana-paper,hyper-paper,umbra-mvcc,oneshot-gc,colibri-paper}) at dedicated servers are not applicable to lakehouses.

Due to these constraints, existing lakehouse systems use a merge-on-read (MoR) approach to replay data changes~\cite{iceberg,hudi,paimon,delta-lake-paper}.
MoR writes the new tuple versions to delta files without changing the base files.
To commit the delta files atomically, the lakehouse executes a checkpoint operation to create a new snapshot (i.e., version) of the table catalog that includes the delta files.
Analytical queries then find the visible snapshot of the table catalog, merge the delta and base files to construct a consistent view of the table.
MoR variants include the LSM-style table layout in Apache Paimon~\cite{paimon-overview}.

\textbf{MoR has major limitations that hinder transactional, real-time data-change replay}:
(1) It has to commit in large batches (e.g., every 5 minutes) because writing delta files and creating a catalog snapshot per mirror transaction is too expensive and would yield unacceptably low replay throughput.
(2) It can not provide cross-table ACID guarantees because catalog snapshot is designed to support atomic file operations on a single table.
(3) It degrades query performance because reading small delta files and merging them with the base data adds high I/O and CPU cost. Background compaction reduces query costs but is itself expensive and may back-pressure replay, as shown in Figure~\ref{fig:compaction-overhead}.

HTAP databases with replicated row-column stores (i.e., dual-layout), such as TiDB~\cite{tidb-paper} and F1 Lightning~\cite{f1-lightning}, also rely on log-based data synchronization.
TiDB can deploy its column store, which is similar to Parquet, to lake storage such as S3.
It replays logs using DeltaTree, a variant of MoR with periodic compaction~\cite{tidb-paper}.
F1 Lightning also applies MoR with compaction to replay data changes on the column store.
Unlike lakehouses, HTAP databases prioritize full transaction guarantees and high freshness over cost efficiency.
\section{Motivation and Design}\label{sec:system-overview}

In this section, we first discuss the opportunities for a more efficient solution for transactional, real-time, and low-cost data-change replay in lakehouses.
Then, we propose our system architecture and core design considerations.

\subsection{Opportunities}\label{subsec:key-observations}

In MoR, the ever-growing tuple versions and small delta files are the root cause of the painful trade-off between data freshness and query performance.
However, in practice, we observe opportunities to eliminate this trade-off without expensive compaction.

\textbf{File combination is more efficient for reducing small reads.}
One main goal of compaction is to reduce small delta-file reads, which are expensive on lake storage~\cite{pixels-s3}, for query execution.
However, to reduce small reads alone, asynchronously rewriting a group of small files into a regular-sized file is more efficient than compaction, because such a small-file combination does not touch the base data~\cite{pixels-hdfs-cache,pixels-s3}.
Our evaluation on AWS S3 shows that 4 EC2 cores can combine small HyBench files of 8192 rows into regular-sized files at 1M row/s, high enough in most scenarios.
With most delta files combined in time, the query I/O pattern amounts to large sequential scans, which is efficient on lake storage.

\begin{figure}[tbp]
	\begin{center}
		{\includegraphics[scale=0.96]{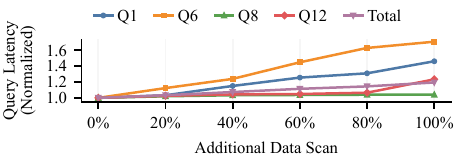}}
	\end{center}
	\caption{\label{fig:scan-extra-data}Query performance with extra data scan on TPC-H.}
    \Description{}
\end{figure}

\textbf{Scanning and filtering out invisible tuple versions is efficient.}
The other goal of compaction is to reclaim stale tuple versions, thus accelerating visibility checks at query runtime.
Workloads in lakehouses are typically dominated by table scans~\cite{mosaic-cache,redshift-redset}, which were assumed to be the main bottleneck for query processing on lake storage~\cite{mosaic-cache,pixels-s3,UnderstandingTablePlacementMothods}.
However, we find this assumption no longer holds in modern lakehouses.
This is because lake storage read bandwidth has improved significantly in recent years~\cite{s3-ec2-bandwidth,pixels-turbo}, and modern columnar file formats are heavily optimized for table scans~\cite{btr-blocks,pixels-s3,ColumnOrdering,AliORC,CarbonDataMainPage,apache-arrow,f3-format}.
In Figure~\ref{fig:scan-extra-data}, we run TPC-H (SF100) queries on AWS S3 to evaluate the scan and filter performance.
The tables are stored in Pixels format, which is optimized for S3 and HDFS~\cite{pixels-s3,pixels-hdfs-cache}.
The queries are run by Trino~\cite{trino-main-page} in an m5.8xlarge EC2 instance.
We randomly duplicate a percentage of tuples to simulate stale tuple versions, which are then read and filtered out during table scan without passing to downstream stages.
The total query latency of all 22 queries increases by only 3.3\%, 7.3\%, 11.4\%, 14.5\%, and 19.4\% when the stale-tuple ratio is 20\%, 40\%, 60\%, 80\%, and 100\%, respectively.
MoR with periodic compaction in Paimon incurs a 39.0\% query slowdown and write stalls in the same environment.
These results indicate that scanning a certain percentage of stale tuple versions and filtering them out during table scan is more efficient.
This observation creates an opportunity to trade a small amount of storage space, the cheapest resource in lakehouses, for higher data change and query performance.

\subsection{Architecture Overview}\label{subsec:architecture-overview}

\begin{figure}[tbp]
	\begin{center}
		{\includegraphics[scale=0.72]{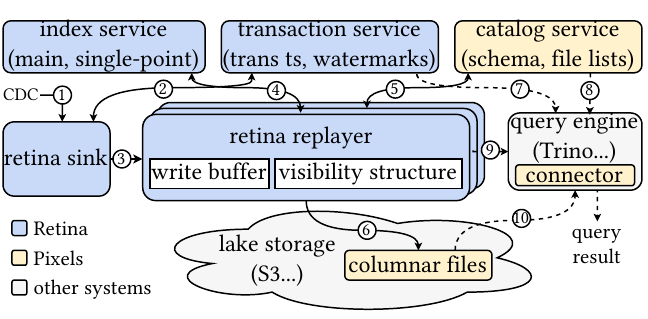}}
	\end{center}
	\caption{\label{fig:retina-architecture}Architectural overview of Retina.}
    \Description{}
\end{figure}

With the observations in Section~\ref{subsec:key-observations}, we propose Retina, a real-time transactional data synchronization framework for lakehouses.
We implement Retina in our open-source lakehouse engine Pixels~\cite{pixels-main-page}, which has a columnar file format optimized for object storage~\cite{pixels-s3} and HDFS~\cite{ColumnOrdering,pixels-hdfs-cache}. 
The Pixels format employs a similar PAX-style~\cite{weaving-relations-for-cache-performance} file layout as Parquet and ORC, and it does not rely on any special features of Pixels.
Thus, migrating Retina to other file formats is practical.

Figure~\ref{fig:retina-architecture} gives an architectural overview of Retina. 
The data-change messages from CDC are consumed by \textit{retina sink} (\textcircled{1}), which allocates transaction timestamps from the \textit{transaction service} (\textcircled{2}) and reconstructs mirror transactions from the data-change messages (Section~\ref{subsec:transaction-reconstruction}).
Inserts, updates, and deletes in the transactions are sharded across a cluster of \textit{retina replayers} (\textcircled{3}).
To replay the data-change operations, the replayer invokes the \textit{index service} to look up and maintain the location of tuples (\textcircled{4}).
Replayer writes new tuple versions to the write buffer and maintains tuple visibility in the visibility structure.
After the replayers successfully replay the data-change operations in a transaction, retina sink requests the transaction service to commit the transaction and advance the high watermark.
The replayer flushes the write buffer into the lake storage using the file ID registered from the catalog service (\textcircled{5} and \textcircled{6}).
The write buffer, visibility structure, and data file management are part of the version storage, which is discussed in Section~\ref{sec:data-storage}.

For an analytical query, the engine obtains the snapshot timestamp from the transaction service (\textcircled{7}) and the schema and file list from the catalog service (\textcircled{8}) during planning. During table scan, the engine workers request the per-row-group visibility bitmap from the retina replayers (\textcircled{9}), which marks whether each tuple version is visible at the query snapshot. Using this bitmap, the reader filters out invisible versions from the row groups read from the columnar files (\textcircled{10}) and returns only visible versions to downstream stages such as joins and aggregations. The write buffer is also read as a special file when it holds tuple versions at or before the query's snapshot timestamp. We detail query processing in Section~\ref{sec:query-processing}.

All the components run in their own processes and communicate using remote procedure calls (RPC), except for the index service, which can optionally run in retina replayers for better performance.
We provide an interactive animation of Retina in its open-source repository~\cite{retina-animation} to better explain how Retina works.

\subsection{Transaction Reconstruction}\label{subsec:transaction-reconstruction}

To support transactional data-change replay, we first reconstruct mirror transactions in retina sink.
The messages extracted by the CDC system are naturally ordered by the commit order of source transactions.
The messages for a source transaction appear consecutively in the CDC stream.
Therefore, we sequentially consume the CDC stream, aggregate messages by transaction ID to form mirror transactions, and assign a unique, monotonically increasing timestamp to each mirror transaction.
The timestamp is allocated by the transaction service.
We do not reuse the timestamps in CDC messages to avoid timestamp collisions between different data sources.
Data-change operations in the transactions are sharded by the primary key.
Operations in a shard (i.e., bucket) are sent to the corresponding retina replayer in small batches.
Retina sink commits the mirror transaction once all its operations are successfully replayed.
We apply group begin and commit to reduce RPC costs.

\subsection{Multi-Version Concurrency Control}\label{subsec:mvcc}

Although lakehouses do not execute user transactions, it is necessary to resolve concurrency conflicts among mirror transactions and analytical queries.
The concurrency control protocol for lakehouses must be cost-efficient, scalable, and comply with the lake storage constraints.
This is very challenging.

\begin{figure}[tbp]
	\begin{center}
		{\includegraphics[scale=0.92]{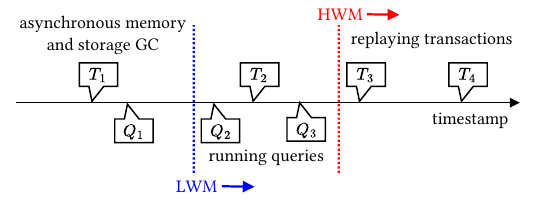}}
	\end{center}
	\caption{\label{fig:retina-watermark}MVCC algorithm in Retina.}
    \Description{}
\end{figure}

Retina proposes the watermark-based MVCC in Figure~\ref{fig:retina-watermark}. 
The high watermark (HWM) means all mirror transactions with earlier timestamps are replayed and visible.
The low watermark (LWM) means all read-only queries with earlier snapshot timestamps have finished.
HWM advances as mirror transactions are committed.
An incoming query takes $(HWM-1)$ as its snapshot timestamp and advances
LWM as it finishes.
A tuple is versioned by the timestamp of the mirror transaction ($create\_ts$) that writes it.
This $create\_ts$ is stored as a hidden column in the columnar files, whereas the $delete\_ts$ of old tuple versions are kept in the visibility structure in retina replayers.
Thus, mirror transactions run at or above HWM, queries run in $[LWM, HWM)$, and asynchronous garbage collection runs below LWM to reclaim memory and storage space.

\textbf{Conflict free.} 
MVCC naturally avoids read-write conflicts (i.e., reads never block later writes).
By letting queries run below $HWM$, they avoid seeing inconsistent states of uncommitted mirror transactions, thereby eliminating write-read conflicts.
For write-write conflicts, the data source ensures source transactions are committed without conflicts; the CDC system ensures source transactions appear in the CDC stream in their commit order; retina sink ensures timestamps are assigned to their mirror transactions in the same order; and retina replayers ensure that changes to the same shard are relayed in the timestamp order.
Mirror transactions therefore can replay in parallel without write-write conflicts.

\textbf{Transaction service.} The transaction service is the centralized authority for timestamps and watermarks in Retina.
The timestamp allocator and the watermarks are atomic integers backed by persistent records in Etcd~\cite{Etcd}, which is highly available.
Thus, the transaction service is fault-tolerant.
With engineering optimizations, it supports beginning and committing over 2 million transactions per second on a single node.

\begin{figure}[tbp]
	\begin{center}
		{\includegraphics[scale=0.72]{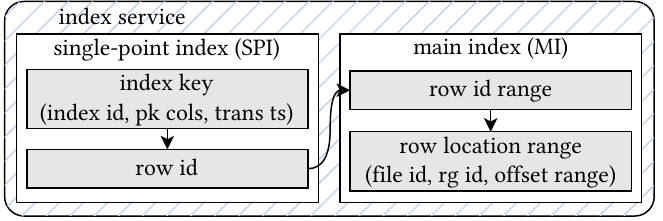}}
	\end{center}
	\caption{\label{fig:retina-index}Index scheme in Retina.}
    \Description{}
\end{figure}

\subsection{Primary Index}\label{subsec:index-service}
To update or delete a tuple, we must locate its primary version in the storage.
Retina has a versioned primary index following the same timestamp ordering rule as the version storage; thus, it requires no additional concurrency control.
As shown in Figure~\ref{fig:retina-index}, the primary index consists of a single-point index (SPI) from index keys to logical row IDs and a main index (MI) from logical row IDs to row locations.
Similar to Colibri~\cite{colibri-paper}, Retina assigns each tuple version a unique, increasing logical row ID.
A row location identifies the file, row group, and row offset of the tuple version.
As tuple versions in a file have consecutive row IDs, MI simply maps row ID ranges to row location ranges.

SPI is versioned: a lookup returns the matching entry with the largest timestamp $\leq$ the request timestamp.
A delete is recorded as a tombstone entry.
Tombstoned SPI entries and the corresponding row ID ranges in MI are purged asynchronously by the garbage collector in the version storage.
The index exposes pluggable interfaces.
In this paper, our implementation backs the SPI with RocksDB~\cite{rocksdb-paper} and the MI with SQLite~\cite{sqlite-paper}.
We allocate a configurable in-memory block cache to accelerate access to RocksDB.
The cache size is 20GB by default and can be tuned according to data-change throughput.
The primary index can be partitioned to scale out.
\section{Version Storage and Data Change}\label{sec:data-storage}

In this section, we introduce the version storage and how data changes and garbage collection are performed on it.

\subsection{Version Storage}\label{subsec:version-storage}

\begin{figure}[tbp]
	\begin{center}
		{\includegraphics[width=0.96\linewidth]{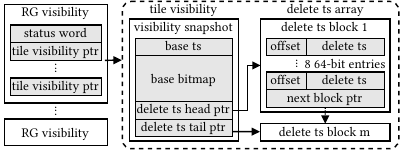}}
	\end{center}
	\caption{\label{fig:visibility}Visibility structure in retina replayer.}
    \Description{}
\end{figure}

The version storage consists of the column store in the lake storage and the auxiliary data structures in the retina replayers.

\textbf{Column store.}
Following the standard storage scheme in lakehouses, a table in Retina is stored as a set of columnar files, each of which contains one or more row groups.
The row groups are assigned to table scan tasks to process in parallel.
To determine whether a tuple version in a row group is visible during table scan, each tuple version should maintain its create timestamp and delete timestamp (if a later version exists).
As the create timestamp does not change after the tuple version is written, we store it as a hidden int64 column in the columnar files, whereas the delete timestamp is maintained in the visibility structure.

\textbf{Write buffer.}
Each table has a write buffer in the replayer to accept new tuple versions.
The write buffer stores data in row blocks that organize data into column vectors.
The write buffer has a single active block that receives new tuple writes and a queue of immutable blocks awaiting background flush to the lake storage.
The write buffer is append-only.
The active block is atomically turned immutable and replaced by a new one when it is full.
Each block has a unique block ID.
The blocks flushed to the lake storage are asynchronously combined into normal-sized columnar files as discussed in Section~\ref{subsec:key-observations}. 
The block size (i.e., the number of tuples per block) is configurable.
Evaluation shows 4096 is sufficient to achieve optimal write performance.
With background flush, each write buffer has 1 active and usually 0-2 immutable blocks, consuming less than 2 MB of memory in our experiments.

\textbf{Visibility structure.}
To determine tuple visibility, we manage the delete marks and delete timestamps of tuple versions in the visibility structure in replayers.
We logically split a row group into tiles, each containing 4096 tuples by default,
and create a \textit{tile visibility} structure for each tile.
All tile visibilities for a row group form a \textit{RG visibility}.
As shown in Figure~\ref{fig:visibility}, tile visibility has a \textit{visibility snapshot} and a \textit{delete ts array}.
The former uses a \textit{base bitmap} to mark whether a tuple in the tile is invisible to queries at or above \textit{base ts}.
The latter is a dynamic array that stores the delete timestamps of tuples deleted after base ts.
It is stored as a linked list of \textit{delete ts blocks} of eight uint64 entries.
The tuple offset in the tile is in the high 12 bits, whereas the tuple's delete timestamp is in the low 52 bits, which are sufficient to allocate timestamps at 1 M/s for 142.8 years.
Each block contains 64 bytes, matching a standard CPU cache line and SIMD vector.
Tuple deletion after base ts appends an entry to the delete ts array.
Queries after base ts can determine tuple visibility by merging delete ts entries into a copy of the base bitmap, accelerated by SIMD.
The delete ts array has no random updates; thus, it is lock-free.
Memory GC merges its head blocks into the base bitmap and releases memory.

\subsection{Data-Change Operations}\label{subsec:data-change-operations}

Retina sink sends data-change operations to the replayer through streaming RPC.
This ensures that data changes arrive at the replayer in the order of their transaction timestamps.

\textbf{Insert operation} provides the after-image of the new tuple.
Retina replayer appends this new tuple to the write buffer and inserts its index entry into the primary index.
Blocks in the write buffer form a virtual file with a single row group and corresponding RG visibility.
A virtual file has a pre-assigned file ID.
When its blocks are combined into a regular file, the file ID is reused.
Thus, row locations stay unchanged, and we do not need to update the primary index when write buffer blocks are flushed or combined.
We create a new virtual file when the current one is full and combined.
As discussed in Section~\ref{subsec:key-observations}, such combination is much more efficient than conventional data compaction.
The same MVCC protocol applies to both regular and virtual files.
Queries can request to read the virtual files, as discussed in detail in Section~\ref{sec:query-processing}.

\textbf{Delete operation} provides the before image of the tuple.
Retina replayer (1) looks up the row location of the tuple's primary version from the primary index, (2) inserts a tombstone into the primary index, (3) finds the corresponding tile visibility by the row location, and (4) appends the transaction timestamp as the delete timestamp into the delete ts array using atomic memory write.
Deletes on the virtual files are no exception.
Queries and deletes do not conflict on the delete ts array because queries run below HWM while deletes replay at or above HWM.
Delete ts entries later than the snapshot timestamp are ignored by the query.

\textbf{Update operation} provides both before-image and after-image of the tuple.
It is a composite operation that deletes the current tuple version and inserts a new one.

\subsection{Garbage Collection}\label{subsec:garbage-collection}

Retina employs asynchronous garbage collection (GC) to reclaim the memory and storage space used by the stale versions that are no longer visible to any running or future queries.

\textbf{Memory GC.}
The growing memory footprint comes from the delete ts in tile visibility.
Since all queries run at or above LWM, versions deleted before LWM are globally invisible.
The garbage collector periodically checks each tile visibility's base ts.
If it falls behind LWM, the garbage collector merges all delete ts blocks below LWM into a copy of the base bitmap, sets the copy's base ts to LWM, and swaps the visibility snapshot pointer atomically before freeing the merged blocks.
This blocks neither data change nor query.

\textbf{Visibility checkpoint.}
Long-running queries may block LWM and memory GC.
To avoid this, Retina has a visibility checkpoint mechanism to detect long-running queries, pre-build and materialize the query's visibility bitmaps to files, force-advance LWM, and resume memory GC.
The checkpoint is only 119 MB per billion tuples; reloading it adds negligible overhead for the long-running query.
Inactive checkpoints are cleaned asynchronously.

\textbf{Storage GC.}
The stale (i.e., globally invisible) tuple versions in the columnar files consume storage space.
Since scanning extra data is inexpensive for analytical queries and the storage is cheap in lakehouses, we perform lazy storage GC.
During memory GC, if we find that the portion of stale tuple versions in a file exceeds a threshold (e.g., 50\%), we push the file ID of the aging file to a pending queue.
Retina replayer launches background storage GC jobs on a daily or weekly basis to rewrite aging files and reclaim storage space.
A storage GC job chooses a safe snapshot timestamp $gts \le LWM$, below all active visibility checkpoints. Picking two aging files at a time, it rewrites the versions visible at $gts$ into a new file.
To ensure file rewrite does not block transactions and queries, we first write the new file and make it visible only to delete operations.
After that, deletes are applied to both new and aging files.
During such dual-delete, the GC job constructs the visibility structure and index entries for new files and synchronizes them with those of the aging files.
When this is done, we make the new file visible and the aging files invisible to later queries atomically in the catalog service.
The GC job will delete the aging files and their visibility structures when no queries are referencing them.
Stale index entries for aging files are purged along with the files.
\section{Query Processing}\label{sec:query-processing}

This section introduces how Retina participates in query processing.

\subsection{An Overview}\label{subsec:query-overview}
We first briefly explain how Pixels interacts with the query engine.
Pixels provides a connector library for the query engine to access the Pixels/Retina services and read the columnar files (Figure~\ref{fig:retina-architecture}).
Using Trino~\cite{trino-main-page} as an example query engine, during query planning, the Trino coordinator invokes the connector to retrieve the schema and file list of each table from the catalog service.
Trino coordinator splits the row groups in the data files into scan splits and assigns them to the Trino workers.
During query execution, each Trino worker launches parallel table-scan tasks to scan the assigned splits using the file reader in the connector.

\subsection{Vectorized-Filter-on-Read}\label{subsec:vfor}

\begin{algorithm}[tbp]
\caption{Vectorized-Filter-on-Read}
\label{alg:vfor}
\KwIn{scan split $s$, query snapshot timestamp $qts$}
\KwResult{visible tuple versions in $s$}

\If{$s$ is a regular file split}{
    \ForEach{row group $rg \in s$}{
        \If{$qts < rg.minCreateTs$}{
            \textbf{continue}\;
        }
        $M \assign \mathrm{GetVisibilityBitmap}(rg,qts)$\;
        \If{$qts < rg.maxCreateTs$}{
            $M \assign M \cap \mathrm{FilterCreateTs}(rg,qts)$\;
        }
        $\mathrm{ReturnTuples}(rg,M)$\;
    }
}
\Else{
    $(ids,rows) \assign \mathrm{ReadVirtualFile}(s,qts)$\;
    $M \assign \mathrm{GetVisibilityBitmap}(s,qts)$\;
    \ForEach{flushed block id $bid \in ids$}{
        $b \assign \mathrm{ReadBlock}(bid)$\;
        $\mathrm{ReturnTuples}(b,M)$\;
    }
    $\mathrm{ReturnTuples}(rows)$\;
}
\end{algorithm}

During query planning, the query engine first obtains a snapshot timestamp (i.e., $HWM-1$) from the transaction service.
This timestamp is passed to table-scan tasks in the scan splits.
Vectorized-filter-on-read (VFoR) then skips invisible tuple versions during table scan and passes each tuple's visible version downstream.
Thus, a query never merges delta and base files or traverses version chains.

As shown in Algorithm~\ref{alg:vfor}, lines 2-8 scan a regular columnar file. 
The file reader first uses the min/max statistics of the hidden $create\_ts$ column in the row group footers to skip row groups whose min $create\_ts$ exceeds the query's snapshot timestamp $qts$ (lines 3-4). 
Then it gets the visibility bitmap $M$ from the replayer (line 5).
If $qts$ is within the min and max $create\_ts$, the reader also filters the row group with the hidden $create\_ts$ column (lines 6-7).
Finally, it uses $M$ as a vectorized mask to filter tuple versions, returning the visible tuple version to the downstream operators (line 8).
Both bitmap-based filtering and timestamp filtering are SIMD-accelerated.
If SQL projections and predicates are pushed down to the file reader, they are also applied during VFoR.

As discussed in Section~\ref{subsec:version-storage}, Retina wraps the write buffer as a virtual file with a single row group.
Lines 10-15 in Algorithm~\ref{alg:vfor} read the virtual file and filter out invisible tuples. The flushed blocks in the virtual file are read from the lake storage (line 13). Only the in-memory blocks are read from the replayer (line 10).
The split for the virtual file is scheduled the same as regular splits. It is dispatched to a single worker in the query cluster. No matter how large the query cluster is, the write buffer is read by a single node. Therefore, the write buffer is not a bottleneck for query scalability.

\subsection{Discussion}\label{subsec:discussion}

\textbf{Isolation level.} 
Retina ensures that mirror transactions are replayed in the commit order of their source transactions, and the query observes a consistent snapshot of committed mirror transactions.
This provides serializable snapshot isolation.
However, if we consider the data source and the lakehouse as a monolithic HTAP system, the global isolation level depends on the data source.

\noindent\textbf{CDC delivery.}
Retina requires that the transaction-grouped data-change messages be delivered to the lakehouse exactly once and in order.
This can be guaranteed by modern CDC systems~\cite{debezium,flink-cdc}.
For CDC setups that do not provide exactly-once and strictly-ordered guarantees, retina sink is responsible for de-duplication and ordering using transaction information in the messages.

\noindent\textbf{Fault recovery.}
The CDC system can persist the recent messages (e.g., in the last 7 days) to a distributed message queue such as Kafka~\cite{kafka-paper}.
This is naturally a write-ahead log (WAL) for Retina.
When a sink or replayer fails, Retina can redo data changes from the latest WAL checkpoint.
The checkpoint is the timestamp and corresponding message offset at which the index, write buffer, and visibility structures of all mirror transactions before it have been saved to persistent storage.
We have implemented this recovery mechanism in Retina.
Unlike user data, states including the timestamp/ID allocators, watermarks, and node heartbeats are backed by Etcd~\cite{Etcd}, a highly available state store.

\noindent\textbf{Point and range query.}
For point queries, the Retina index service also supports secondary indexes.
In non-unique secondary indexes, we store the row ID as the postfix of the SPI index key, and queries search SPI using a prefix without a row ID.
Range queries are generally supported by data skipping and range partitioning in lakehouses.
Pixels supports data skipping.
Range partitioning can be supported by using a range-based sharding policy in retina sink.
\section{Evaluation}\label{sec:evaluation}

\subsection{Experimental Setup}\label{subsec:setup}

\textbf{Systems to Evaluate.}
We evaluate Retina against four state-of-the-art systems: Apache Iceberg~\cite{iceberg}, Apache Paimon~\cite{paimon}, TiDB~\cite{tidb-paper}, and Colibri~\cite{colibri-paper}.
Iceberg is a popular lakehouse engine that employs a MoR approach for data changes.
The table format in Iceberg is the de facto standard for today's lakehouses.
We evaluate Iceberg v2.
Paimon is a lakehouse engine designed for real-time analytics.
It supports fine-grained MoR that partitions tables into buckets and stores each bucket as a columnar LSM-Tree. 
It has an optimization named the deletion vector (DV) to accelerate MoR.
Similar to VFoR, DV can mark tuples in a file as deleted.
However, it does not provide snapshot isolation for queries, and it is generated during compaction; thus, it may further degrade data-change performance and data freshness.
We evaluate Paimon with DV enabled or disabled.
Both Iceberg and Paimon store data in Parquet format.
They run as embedded SDKs in compute systems such as Flink~\cite{flink-paper}.
TiDB is a representative HTAP database system with a dual-layout architecture.
It stores data in row format in its OLTP cluster, TiKV, and in columnar format in its OLAP cluster, TiFlash.
Data changes in TiKV are replicated to TiFlash via RAFT~\cite{raft-paper} logs.
This is similar to log-based CDC in lakehouses. 
TiFlash can store data in a columnar format similar to Parquet in lake storage~\cite{tidb-paper}.
Colibri is an academic HTAP database system originating from Umbra~\cite{umbra-paper}.
It can store base data in the hybrid Data Block format~\cite{data-blocks-tum} in object storage.

\noindent\textbf{Compute and Storage.}
All systems run on AWS EC2 \texttt{i7i.4xlarge} instances, each with 16 vCPUs, 128~GB RAM, 3.75~TB local SSD, and up to 25~Gbps network.
Cluster size varies across experiments.
All systems store base data in AWS S3 in the same availability zone as the EC2 instances.
Local SSD holds the per-system hot state: Retina's primary index,
the state-backend data of Iceberg and Paimon, TiDB's TiKV row store, and Colibri's WAL and pages.

\noindent\textbf{Benchmarks.}
We use \textit{HyBench}~\cite{hybench} and \textit{CH-benCHmark}~\cite{chbenchmark}, two widely used HTAP benchmarks~\cite{colibri-paper,tidb-chbenchmark,vegito-htap-paper}.
To evaluate data-change performance and freshness, we exclude read-only OLTP transactions, as they do not emit data changes.
Then, the OLTP operations in both benchmarks are dominated by random updates ($\approx70\%$) and inserts ($\approx25\%$).
However, updates in HyBench occur only on small account tables, whereas updates in CH-benCHmark occur primarily on the two largest tables (\texttt{ORDERLINE} and \texttt{STOCK}).
We stream data changes to the systems after loading the base datasets.
We measure the row throughput in this paper, as it is a common metric for CDC and lakehouses~\cite{debezium-throughput,paimon-iceberg-throughput}.
The average number of rows changed per transaction is 10.23 in HyBench and 13.58 in CH-benchmark.

\noindent
\textbf{Baselines.}
The lakehouse systems (i.e., Retina, Iceberg, and Paimon) ingest data changes from the CDC stream.
Thus, they do not pay the costs of OLTP processing.
The HTAP systems (i.e., TiDB and Colibri) do not expose a CDC ingestion interface; hence, we have to submit OLTP transactions to their SQL interfaces.
This introduces additional SQL parsing and OLTP costs.
Therefore, it is not completely fair to directly compare the performance numbers of HTAP and Lakehouse systems.
For OLAP queries, Retina, Iceberg, and Paimon use a Trino cluster with one coordinator and five workers unless stated otherwise; TiDB and Colibri use their native engines.
We use the recommended settings for all systems.
Iceberg uses upserts to reduce write latency.
Paimon uses a relaxed compaction trigger to reduce compaction pressure during writes.
TiDB creates the indexes required by the workload.
We run Iceberg and Paimon in Flink.
The catalog checkpoint interval in Flink is varied in the freshness experiment and fixed at 3~min elsewhere.
In the single-node experiments, Retina uses one replayer; Iceberg and Paimon each use one Flink node; and Colibri uses one instance.
TiDB uses its minimum HTAP deployment of three TiKV nodes and one TiFlash node except in Section~\ref{subsec:scalability}.

\subsection{Freshness and Throughput}\label{subsec:freshness}

\begin{figure*}[tbp]
	\begin{center}
		{\includegraphics[scale=0.96]{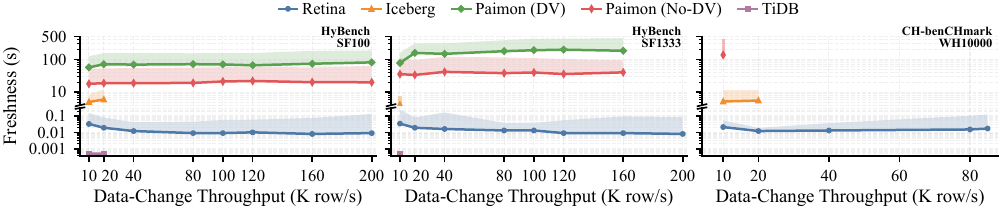}}
	\end{center}
	\caption{\label{fig:freshness} Freshness vs. throughput. Solid lines represent median (P50) freshness, shaded regions represent P99 tail freshness.}
    \Description{}
\end{figure*}

In this experiment, we evaluate the impact of data-change throughput on data freshness in a single EC2 instance.
We use the HyBench method to measure data freshness:
each table carries a $fresh\_ts$ column set at transaction time, a query started at $query\_ts$ reads $\max(fresh\_ts)$, and freshness is $query\_ts-\max(fresh\_ts)$.

For Iceberg and Paimon, freshness depends on the catalog checkpoint interval, since replayed changes become visible after checkpointing. We therefore vary it in this experiment.
As shown in Figure ~\ref{fig:freshness}, Retina delivers consistent 10-millisecond-level median data freshness regardless of data-change throughput.
On HyBench, the P99 freshness remains below 250 ms.
On CH-benCHmark,
Retina sustains 80K row/s throughput with a P99 freshness of 103 ms.
The freshness is slightly worse at low throughput due to the measurement method.
Freshness measurement queries are executed periodically.
At low throughput, the time interval between adjacent write transactions is long, leading to an overestimation of freshness.

Apache Iceberg achieves good freshness only at low throughput. On HyBench SF1333, it achieves a P99 freshness of 7.66s at 10K row/s but fails to scale further.
The failure stems from I/O latency on object storage.
High-frequency commits generate an explosion of small files and catalog snapshots. The I/O bottleneck of writing small files back-pressures the data-change pipeline, causing write stalls.
Iceberg performs similarly on CH-benCHmark WH10000, as its simple MoR without compaction is not sensitive to heavy updates on large tables.
Apache Paimon sustains high throughput but sacrifices freshness.
Paimon (No-DV) exhibits better freshness than Paimon (DV), because DV maintenance introduces significant overhead for the write pipeline.
Paimon (DV) fails to run on CH-benCHmark WH10000.
Paimon (No-DV) also performs much worse on this workload.
This is due to the expense of frequent LSM compaction in the table buckets, which makes Paimon hard to handle heavy updates on large tables. DV maintenance makes compaction performance even worse in this case.

For TiDB, we consulted the TiDB tech team to tune it at best effort.
TiDB fails to sustain high throughput. The Raft consensus protocol is the main bottleneck limiting transaction throughput.
Compared with lakehouse systems, TiDB excels at providing row-level transaction guarantees and better freshness (0.5ms P50 and P99 freshness on HyBench).
The data freshness of TiDB is measured by its own performance metric \textit{raft write log apply duration}.
Colibri can only run on HyBench SF100, providing 4K row/s throughput, 545.5ms P50 freshness, and 1092ms P99 freshness, with 16 TP clients.
This may be because Colibri is optimized for hotspot updates~\cite{colibri-paper}, whereas updates in HyBench are more random.

\textbf{Summary:} Retina is the only system that achieves 10-ms-level freshness under high data-change throughput.

\subsection{Query Performance Consistency}\label{subsec:query_perf}

\begin{figure}[tbp]
	\begin{center}
        {\includegraphics[scale=0.96]{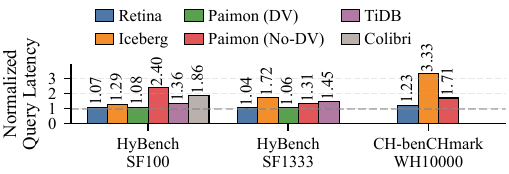}}
	\end{center}
	\caption{\label{fig:ap-geo-mean}Normalized geometric mean of OLAP query latency under 1\% delta data and 10K row/s data-change throughput.}
\end{figure}

In this experiment, we evaluate the impact of data changes on query performance.
Figure~\ref{fig:ap-geo-mean} shows the normalized geometric mean of OLAP query latency under a background data-change workload with 1\% delta data.
Each value is normalized to the corresponding system's query latency on static base data.
The data-change rate is 10K row/s except for Colibri, which caps at 4K row/s.

Retina demonstrates superior performance consistency across all benchmarks, especially on CH-benCHmark, where heavy updates generate many stale versions.
This validates the efficiency of our VFoR design.
To further evaluate whether this robustness holds under different data-change throughput, we add a separate experiment on HyBench SF1333.
We increase the data-change throughput from 10K to 160K row/s, which only Retina and Paimon (No-DV) can sustain on a single node. The normalized query latency for Retina and Paimon (No-DV) remains around 1.04 and 1.33, respectively.

Apache Iceberg exhibits poor query performance on the large datasets.
Queries suffer from the many small files and catalog snapshots it produces.
Paimon (No-DV) suffers from the I/O amplification of its LSM-Tree structure.
Queries must perform expensive merge sorting on numerous small files during execution.
Paimon (DV) exhibits consistent performance on HyBench because HyBench updates only small account tables while its large tables receive mostly inserts, so deletion vectors stay small and avoid most merge overhead. On CH-benCHmark WH10000, where updates fall on large tables and generate many versions, Paimon (DV) fails even at 1K row/s.
TiDB also fails to execute the queries on CH-benCHmark WH10000.
Colibri can only run on HyBench SF100 and exhibits high query performance degradation.

\textbf{Summary:} Retina provides the best query performance consistency under heavy data-change workloads.

\subsection{Scalability}\label{subsec:scalability}

\begin{figure}[tbp]
	\begin{center}
		{\includegraphics[scale=0.96]{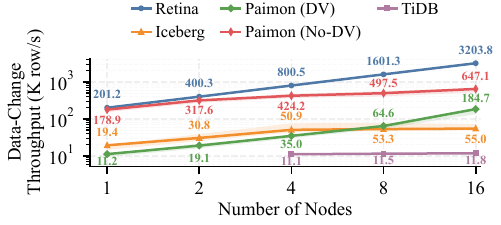}}
	\end{center}
	\caption{\label{fig:scale}Maximum sustainable data-change throughput under varying cluster size on HyBench SF1333.}
\end{figure}

We measure horizontal scalability by scaling out each system's data-change components and recording the maximum sustainable data-change throughput on HyBench SF1333. 
For each cluster size, we run 30-minute trials at increasing throughputs and keep the highest stable rate.
Colibri is single-node and excluded in this experiment.
For TiDB, we use (TiKV, TiFlash) = (3,1), (6,2), and (12,4) for 4, 8, and 16-node clusters, respectively, as recommended.

As shown in Figure ~\ref{fig:scale}, Retina demonstrates linear scalability.
With 16 nodes, the throughput reaches 3.2M row/s. 
This is thanks to the lightweight MVCC mechanism and efficient version storage.
Index operations become the main bottleneck during replay.
Hence, we scale and co-locate the index service with the replayers.
Apache Iceberg does not scale after 4 nodes.
This is because committing catalog snapshots becomes the main bottleneck under high concurrency.
Paimon (No-DV) also exhibits poor scalability. Although its LSM-Tree design is write-friendly, the massive number of small files triggers S3 request throttling, making it I/O-bounded under high concurrency.
Paimon (DV) demonstrates the illusion of linear scalability.
This is because maintaining DVs during data compaction incurs substantial memory overhead, and data-change performance is memory-bound on small clusters.
The throughput of TiDB plateaus at 11-12K row/s. This is because the heavy updates on the small tables target only a few ranges on a few nodes.
The load balancer attempts to ship these hot ranges across the cluster, but this does not help improve throughput.

Furthermore, Retina maintains superior throughput stability as it scales to large clusters.
The standard deviation (SD) of its throughput remains below 2\% at 16 nodes.
The baselines show significant throughput fluctuations (e.g., an SD of 111k at 16 nodes for Paimon No-DV) due to background compaction and write conflicts.

\textbf{Summary:} Retina is the only system that scales linearly to 3.2M row/s stable data-change throughput.

\subsection{Resource Efficiency}\label{subsec:resource}

In this experiment, we evaluate CPU and memory consumption under a stable data-change throughput on a single node using HyBench and CH-benCHmark.
The results are shown in Table~\ref{tab:resource_usage}.

\begin{table}[t]
    \centering
    \caption{Mean resource usage under 10K row/s data-change.}
    \label{tab:resource_usage}
    \begin{adjustbox}{max width=\columnwidth}
    \begin{tabular}{lcccc}
        \toprule
        \textbf{System} 
        & \multicolumn{2}{c}{\textbf{HyBench SF1333}} 
        & \multicolumn{2}{c}{\textbf{CH-benCH WH10000}} \\
        \cmidrule(lr){2-3} \cmidrule(lr){4-5}
        & \textbf{CPU (\%)} & \textbf{Mem (GB)}
        & \textbf{CPU (\%)} & \textbf{Mem (GB)} \\
        \midrule
        Iceberg & 183.05 & 39.57 & \textbf{128.04} & 42.19 \\
        Paimon (No-DV) & 817.30 & 25.51 & N/A* & N/A* \\
        Paimon (DV) & 581.71 & 27.67 & N/A* & N/A* \\
        TiDB & 2393.79 & 84.30 & N/A* & N/A* \\
        Retina & \textbf{62.17} & \textbf{19.28} & 644.13 & \textbf{41.34} \\
        \bottomrule
    \end{tabular}
    \end{adjustbox}
    \footnotesize
    \textit{*} N/A indicates the system failed to load or run the benchmark.
\end{table}

On HyBench SF1333, Retina consumes only 62.17\% CPU and 19.28 GB of memory.
Paimon consumes 581.71\% or 817.3\% CPU with or without DV, due to frequent background compaction required by the LSM-tree architecture.
Iceberg exhibits lower CPU usage at 183.05\% but requires 39.57 GB of memory.
This stems from the managed memory allocation in its Flink~\cite{flink-paper} and RocksDB~\cite{rocksdb-paper} state backends.
The resource usage of TiDB includes both TiKV and TiFlash.
On CH-benCHmark WH10000, Retina incurs higher CPU usage due to heavy updates on large tables,
which trigger frequent seek-and-read operations to retrieve the latest visible index entry from the RocksDB-backed single-point index (SPI).
In RocksDB, this is more expensive than point operations on very large datasets.
This overhead can be mitigated by partitioning the SPI across more RocksDB column families.
Since our indexing framework provides a pluggable interface for index implementations, this issue can be addressed by replacing the index backend with a more efficient one.
We also run Colibri with 16 TP clients on HyBench SF100.
It consumes 1595.53\% CPU and 84.24 GB of memory on average when processing data changes at 4K row/s.

To understand how memory consumption scales with data size, we also perform the experiment on HyBench SF100.
From SF100 to SF1333, memory increases from 26.82 GB to 39.57 GB for Iceberg.
The larger table increases the states maintained by Flink, leading to a larger runtime memory footprint.
Paimon's memory usage is relatively stable, increasing from 23.20 GB to 25.51 GB without DV and remaining almost unchanged at 27.70 GB and 27.67 GB with DV.
This is because Paimon has bounded-size memory tables for the LSM trees.
Retina increases from 12.40 GB to 19.28 GB.
This increase mainly comes from the growing block cache in the RocksDB index backend, rather than the visibility structure.
We can bound this block cache by slightly compromising data-change throughput.

\begin{figure}
	\begin{center}
		{\includegraphics[scale=0.96]{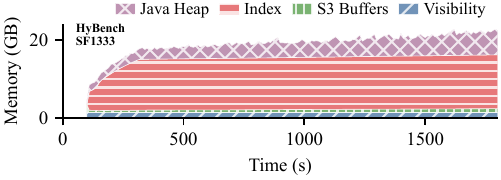}}
	\end{center}
    \caption{\label{fig:retina_resource}Memory breakdown of Retina on HyBench SF1333.}
\end{figure}

As shown in Figure ~\ref{fig:retina_resource}, we further break down the memory consumption of Retina on HyBench SF1333 under 10K row/s data-change throughput.
The visibility structure is remarkably lightweight.
Its memory footprint is only 1.56 GB, i.e., 1.16 bit/tuple given the 11.6 billion tuples in HyBench SF1333.
This footprint remains stable over time, confirming the effectiveness of memory GC.
Most memory is used by the index and the Java heap.
RocksDB warms up a large block cache to accelerate lookups.
The growing Java heap is due to the lazy memory-release mechanism in JVM.
S3 buffer is allocated by the AWS SDK.
The write buffer has a memory footprint below 2 MB per table per replayer, which is negligible.
The memory breakdown on CH-benCHmark WH10000 is similar, with visibility consuming 1.97 GB under the heavy updates.

\begin{figure}
	\begin{center}
		{\includegraphics[scale=0.96]{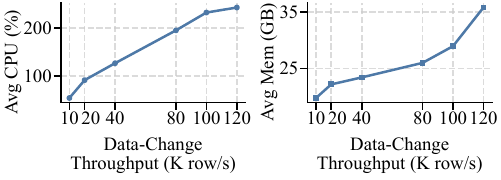}}
	\end{center}
	\caption{\label{fig:retina_resource_scale} Resource usage of Retina under varying data-change throughput on HyBench SF1333.}
\end{figure}

As shown in Figure~\ref{fig:retina_resource_scale}, as the data-change throughput increases on HyBench SF1333, Retina achieves near-linear CPU usage with an R2 value of 0.98. The marginal cost is only 1.73\% CPU per 1K row/s.
CH-benCHmark shows higher absolute resource consumption due to complex update patterns. However, Retina still maintains near-linear scaling, at 3.29\% CPU per 1K row/s.

The S3 storage overhead of Retina mainly comes from the unreclaimed stale tuple versions and the hidden timestamp column.
With storage GC, stale tuple versions introduce less than 2.4\% storage overhead in HyBench evaluations.
The hidden column can be compressed effectively as the create timestamps in a file are naturally ordered.
For data files bulk-loaded with the same create ts, the storage overhead of the hidden column is 0.18 bit/tuple.
For data files created during replay, the storage overhead of the hidden column is 1.85 bit/tuple (1.29\% of space) for HyBench and 7.94 bit/tuple (0.73\% of space) for CH-benCHmark.
The storage overhead of Iceberg mainly comes from the delta files; hence, it increases with the ratio of delta data.
Whereas for Paimon (DV and No-DV), the storage overhead on HyBench can reach 70\%-110\% mainly due to the storage-space amplification in LSM trees.

\textbf{Summary:} The resource usage of Retina scales linearly with data-change throughput and is lower than that of the baselines in most conditions. The consumption primarily stems from the single-point index backend, RocksDB, which is a replaceable component.

\subsection{Effects of Tuning Knobs}\label{subsec:ablation}

In this experiment, we evaluate the effects of critical tuning knobs in Retina.
The evaluations are performed on HyBench SF1333.
Data changes are processed on a single node at a data-change throughput of 200K row/s, and queries are executed in a separate Trino cluster with 1 coordinator and 5 workers.

\begin{figure}[t]
	\centering
	\begin{subfigure}[b]{0.48\columnwidth}
		\centering
		\includegraphics[width=\linewidth]{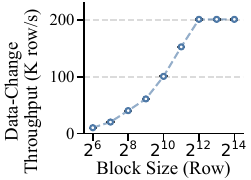}
		\caption{Data-Change throughput.}
		\label{fig:memtableSize:throughput}
	\end{subfigure}
	\begin{subfigure}[b]{0.48\columnwidth}
		\centering
		\includegraphics[width=\linewidth]{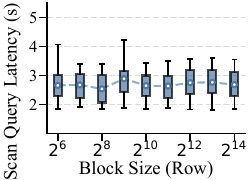}
		\caption{Scan query latency.}
		\label{fig:memtableSize:latency}
	\end{subfigure}
	\caption{\label{fig:memtableSize} Effects of write buffer block size.}
    \Description{}
\end{figure}

\textbf{Write buffer block size.} Figure ~\ref{fig:memtableSize} shows the effect of write buffer block size on data-change throughput and scan performance. We use a simple table-scan query to evaluate the scan performance.
We can see that data-change throughput increases with block size. Larger blocks reduce the frequency of block flushes and amortize the cost of metadata updates. With 4 background block flush threads, data-change throughput saturates at a block size of 4096 rows. At this point, the bottleneck shifts from the write buffer to other replay operations.
The replayer must deserialize each incoming data-change request and update the primary index. Such per-tuple costs remain constant regardless of block size. Scan performance remains stable across varying block sizes.
As discussed in Section~\ref{subsec:vfor}, the query engine treats the write buffer as virtual files and applies the same VFoR approach to scan them.
This improves the scan efficiency and ensures it is not affected by the write buffer block size.
We use a default block size of 8192 rows in Retina.

\begin{figure}[t]
	\begin{center}
		{\includegraphics[scale=0.96]{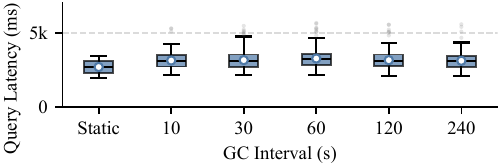}}
	\end{center}
	\caption{\label{fig:gc-interval} Effects of GC interval.}
\end{figure}

\textbf{GC interval.} Figure ~\ref{fig:gc-interval} shows the effect of memory GC intervals on scan performance.
We compare active data change with GC against a static-data baseline.
The static baseline has a median scan query latency of 2657 ms.
Under active GC, the query latency remains stable at approximately 3100 ms across all intervals from 10 to 240 seconds. The performance overhead remains constant at approximately 450 ms regardless of GC frequency. This proves that the lock-free visibility structure in Retina successfully isolates analytical queries from background maintenance tasks.
We use a default memory GC interval of 60 seconds in Retina.

\textbf{Combination granularity.} We evaluate the effect of block combination granularity, i.e., the number of write buffer blocks combined into a regular column file.
The results show that it has no obvious effect on data freshness or query performance.

\begin{figure}[t]
	\begin{center}
		{\includegraphics[scale=0.96]{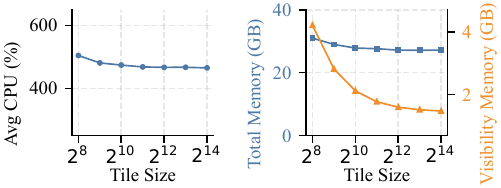}}
	\end{center}
	\caption{\label{fig:tileSize} Effects of visibility tile size.}
\end{figure}

\textbf{Tile size.}
Figure~\ref{fig:tileSize} shows the effects of visibility tile size.
We observe that tile size primarily affects the memory usage of visibility structures.
Each tile visibility structure has an inevitable memory overhead for the delete ts array, base ts, and tile visibility ptr.
A smaller tile size leads to more tile visibilities and higher memory usage.
Such memory usage for visibility approaches the theoretical lower bound of 1 bit per tuple version when the tile size increases.
Evaluations also show that query and data-change performance are not affected by tile size.
We use 4096 as the default tile size in Retina.
It provides favorable memory efficiency and requires 12 bits to store the tuple offset in the tile. This leaves enough (i.e., 52) bits within the delete ts entry to store the delete timestamp.

\subsection{Ablation Study}\label{subsec:component} 
In this section, we perform an ablation study to quantify the contributions of the Retina components from different dimensions.
Since Section~\ref{subsec:resource} has analyzed the resource consumption of the components, we focus on transaction guarantees, data-change throughput, data freshness, and query performance in this section.

\textbf{Transaction guarantees.}
The catalog snapshot mechanism in existing lakehouses, in fact, implements a git-style serial history for table catalogs rather than a full MVCC for user data~\cite{lake-tables-short}.
It does not support cross-table transactions, row-level ACID, or concurrent writes~\cite{lake-tables-short}.
The lightweight MVCC protocol and the corresponding version storage in Retina enable it to replay transactions from OLTP sources and provide full ACID guarantees.

\textbf{Data-change throughput.}
Retina can not execute end-to-end data changes if we ablate any essential components. Therefore, we performed unit tests on each component on a single i7i.4xlarge instance to understand its impact on data-change throughput.
Results show that: 
(1) The transaction service can begin and commit 2.67M transactions/s.
This is equivalent to 27.31M and 36.26M row/s on HyBench and CH-benCHmark, respectively.
(2) Each write buffer can accept 3.37M inserts/s.
(3) The visibility structure can process 72.7M deletes/s.
(4) A single index service instance can sustain up to 253K updates/s.
We can see that the index throughput is close to the 200K row/s max end-to-end data-change throughput on a single node (see Figure~\ref{fig:freshness}), while the throughput of the other components is 1-2 orders of magnitude higher.
In other words, the index service is not only the main resource consumer (see Section~\ref{subsec:resource}), but also the main throughput bottleneck.
However, indexes are necessary for efficient data updates.
The deletion vector in Paimon also requires a primary index to locate the tuple versions.
Paimon's way of storing a table as LSM-Tree buckets provides a form of clustered primary index~\cite{lake-tables-short}.
However, such a storage layout incurs other bottlenecks as shown in Sections~\ref{subsec:freshness} and~\ref{subsec:scalability}.

\begin{figure}[t]
	\begin{center}
		{\includegraphics[scale=0.96]{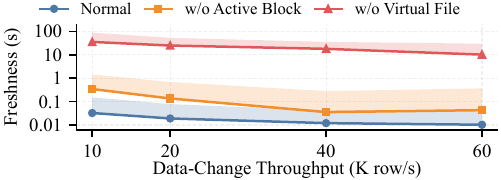}}
	\end{center}
    \caption{\label{fig:freshness_ablation}Data freshness under different visibility settings.}
\end{figure}

\textbf{Data freshness.}
The superior data freshness of Retina primarily stems from its novel visibility mechanism.
This is more than just allowing queries to read the write buffer; it also requires the appropriate visibility structures and MVCC protocol to ensure data consistency.
However, to study its effects on data freshness, as shown in Figure~\ref{fig:freshness_ablation}, we force the queries not to read the active block in the write buffer (w/o Active Block) or the virtual files (w/o Virtual File) at the risk of violating data consistency.
The result confirms that this ability is essential to achieve good data freshness.

\textbf{Query performance.}
Section~\ref{subsec:query_perf} demonstrates that Retina has lower query overhead compared to the baselines.
In this experiment, we further evaluate the overhead of VFoR compared to the ideal situation of disabling visibility check (i.e., completely not accessing replayers in the queries).
The result shows that on HyBench SF100 and a single node, enabling VFoR increases query latency by only 2.33\%.
This confirms VFoR is close to an optimal visibility solution.
\section{Related Work}\label{sec:related-work}

\textbf{HTAP database systems} typically have different architectures and application scenarios than real-time lakehouses; however, they share the common goal of improving data freshness for analytical workloads and minimizing performance interference between transaction processing and query execution.
In-memory HTAP databases, such as HyPer~\cite{hyper-paper}, VEGITO~\cite{vegito-htap-paper}, HYRISE~\cite{hyrise}, and Proteus~\cite{proteus-htap}, achieve this goal through dedicated in-memory storage layouts and concurrency-control algorithms.
They assume that a majority of the dataset fits in memory, making them suitable for applications with small datasets but extremely high performance requirements.
Some other HTAP databases, such as SAP HANA~\cite{hana-paper}, Umbra~\cite{umbra-paper}, and OceanBase~\cite{oceanbase-htap}, extend their storage engines to high-performance external storage, such as direct-attached SSDs.
These types of HTAP databases differ fundamentally from lakehouses in their underlying storage, leading to significant differences in architectural design.
However, with the wide usage of cloud object storage in OLAP systems~\cite{snowflake-nsdi20,delta-lake-paper,pixels-s3}, some HTAP databases, such as Colibri~\cite{colibri-paper} and TiDB~\cite{tidb-paper}, begin to bridge the gap by supporting object storage, which is the common underlying storage in lakehouses, as an optional external storage system.
The design considerations in this type of HTAP database offer valuable insights for lakehouses.
For example, the row ID allocation and main index schemes in Colibri are inspiring for the index design in Retina.

\textbf{Timestamp watermarks} are commonly used to implement MVCC in relational databases.
For example, the MVCC implementation in HyPer~\cite{hyper-paper}, Hekaton~\cite{hekaton-paper}, SAP HANA~\cite{hana-paper}, and Deuteronomy~\cite{deuteronomy-paper} use a watermark to drive garbage collection (GC).
Similar to the low watermark in Retina, transactions in these database systems run above the watermark; thus, stale tuple versions in the version chain that fall behind the watermark can be garbage-collected.
This is considered lazy GC in~\cite{steam-gc}, which proposes an approach to eagerly collect stale versions above the watermark.
In contrast, we propose the visibility checkpoint approach that can push the low watermark over the snapshot timestamp of long-running queries.
This provides favorable GC efficiency while enabling a simpler and more scalable architecture, which is important for lakehouses.
Besides, C-Store~\cite{cstore-stonebraker2005} has an epoch-based watermark that enables snapshot isolation, similar to the epoch-based snapshot isolation in VEGITO~\cite{vegito-htap-paper}.
This is inspiring for the MVCC protocol design in Retina.
The main contribution of this paper is to propose a lightweight and scalable MVCC mechanism tailored for lakehouses, rather than the watermarks themselves.

\textbf{Deletion vectors/bitmaps} are widely used to accelerate filtering of deleted tuples in column stores and databases such as Paimon~\cite{paimon}, Iceberg v3~\cite{iceberg-spec}, StarRocks~\cite{starrocks}, and Colibri~\cite{colibri-paper}.
However, deletion bitmaps alone are insufficient for a version storage to determine whether a tuple version is visible to an active query.
Therefore, these systems either forgo full transaction support and only support file-level catalog version control (e.g., Paimon, Iceberg v3, and StarRocks) or require additional auxiliary data structures, such as the version chain in Colibri, to support snapshot isolation.
In this paper, we propose a novel version storage that supports snapshot isolation while achieving a high memory efficiency close to that of bitmaps. It provides superior performance, scalability, and resource efficiency for lakehouses.
In addition to deletion bitmap, Iceberg v2~\cite{iceberg-spec} and Vertica~\cite{vertica-paper} use \textit{positional delete file/vector} to maintain the locations of deleted tuples, whereas C-Store uses \textit{deleted record vector} to maintain the epoch in which the tuple was deleted.
These approaches are less efficient than deletion bitmap.
\section{Conclusion}\label{sec:conclusion}

This paper proposes a novel data synchronization framework, Retina, for lakehouses.
It supports efficient data-change operations without expensive version merging and data compaction, based on the observation that a small amount of additional data scanning and storage usage incurs negligible performance and monetary overheads for modern lakehouses.
By making minor, beneficial trade-offs in storage space and sequential reads, we design the lightweight MVCC algorithm and corresponding versioned storage tailored for real-time analytics in lakehouses.
Evaluations show that Retina provides 10-ms-level data freshness, over 3.2M row/s scalable replay throughput, and transaction guarantees, without compromising query performance or cost efficiency, significantly outperforming state-of-the-art lakehouses and lake-compatible HTAP databases.

\begin{acks}
We thank Professor Anastasia Ailamaki (EPFL) for her constructive suggestions and feedback that significantly improved the quality of this research.
We also thank Zhipeng Mao and Tianxiao Shen for their participation in an early-version implementation of Retina during their master's studies at EPFL.
That version provided valuable experience for the final design and implementation of Retina.
\end{acks}

%\clearpage

\bibliographystyle{ACM-Reference-Format}
\bibliography{my}

\end{document}